\documentclass{article}
\usepackage{amsmath}
\usepackage{amssymb}
\usepackage{graphicx}
\usepackage{subcaption} 
\usepackage[a4paper, left=2.5cm, right=2cm, top=2cm, bottom=3cm]{geometry}
\usepackage{url} 
\usepackage[breaklinks=true]{hyperref} 

\title{Demographic Dynamics and Artificial Intelligence: Challenges and Opportunities in Europe and Africa for 2050}
\author{Mohamed El Louadi\\
University of Tunis\\
Institut Supérieur de Gestion\\
41 rue de la Liberté – Cité Bouchoucha\\
2000 Le Bardo -- Tunisia\\
\texttt{mohamed.louadi@isg.rnu.tn}}
\date{}

\begin{document}

\maketitle

\begin{abstract}
This paper explores the complex relationship between demographics and artificial intelligence (AI) advances in Europe and Africa, projecting into the year 2050. The advancement of AI technologies has occurred at diverse rates, with Africa lagging behind Europe. Moreover, the imminent economic consequences of demographic shifts require a more careful examination of immigration patterns, with Africa emerging as a viable labor pool for European countries. However, within these dynamics, questions are raised about the differences in AI proficiency between African immigrants and Europeans by 2050. This paper examines demographic trends and AI developments to unravel insights into the multifaceted challenges and opportunities that lie ahead in the realms of technology, the economy, and society as we look ahead to 2050.
\end{abstract}


\section{Introduction}
The confluence of demographic shifts and technological advancements is a distinctive characteristic of the global landscape between Africa and Europe. The nuanced interplay between demographic trends and the evolution of artificial intelligence (AI) has profound implications for societies and economies worldwide. Africa and Europe are focal points marked by distinctive demographic profiles and developmental trajectories.

Africa emerges as a locus of youthful dynamism, harboring the world’s youngest population cohort. This demographic dividend, characterized by a burgeoning youth demographic, portends transformative potential across various spheres of human endeavor. Conversely, Europe grapples with the ramifications of demographic aging, navigating the complexities inherent in managing an increasingly elderly population within existing socio-economic frameworks.

Simultaneously, the advent of AI technologies engenders transformative possibilities and challenges, permeating sectors ranging from healthcare and finance to governance and education. Notwithstanding its transformative potential, the diffusion of AI technologies unfolds unevenly across regions, with Africa confronting notable disparities in technological adoption and innovation compared to its European, American, and Asian counterparts.

Moreover, the intersection of demographic shifts and AI development underscores the need to address disparities in technological competencies, particularly as African immigrants navigate economic opportunities in Europe.

Thus, against the backdrop of demographic diversity and technological innovation, this paper explores the connection between demographic changes and technological advancement. Its goal is to provide useful insights to scholars, policymakers, and stakeholders interested in shaping our future.

It explores the intricate relationship between demographic shifts and advancements in artificial intelligence (AI) in Europe and Africa. It delves into the disparities in AI development rates between the two continents, with Africa lagging behind Europe. It also examines the economic implications of demographic changes, particularly in relation to immigration patterns, highlighting Africa as a potential labor pool for European countries.

In the next section, we delve into recent advancements in AI technologies, focusing on their potential impact on future developments. In the following section, we examine demographic shifts, highlighting Europe's aging demographics and imminent dependence on immigration. We also focus on Africa, discussing its demographic dividend characterized by a burgeoning youth demographic. We explore the opportunities and challenges presented by Africa's demographic profile and their implications for the continent's future development. In the third section, we delve into the intersection of Africa, Information and Communication Technology (ICT), and AI. We discuss the current landscape of technological adoption in Africa and the potential of AI to drive innovation and economic growth. In the fourth section, we address the challenges posed by the lack of digital culture in Africa. We explore how overcoming this barrier is crucial for realizing the full potential of the data economy and fostering technological advancements. In the fifth section, we examine the evolving landscape of work in the era of AI, discuss predictions about job displacement and creation, and discuss how AI is expected to impact labor productivity and job markets in Europe and Africa.

We conclude by summarizing the key insights presented in this paper. The conclusion reflects on the challenges and opportunities discussed in the preceding sections and offers implications for policymakers, scholars, and stakeholders interested in shaping the future of technology, the economy, and society in Africa and Europe.

\section{AI and its development}
The recent increase in the popularity of AI has been propelled by notable advancements in AI technologies and their increasing integration into everyday life. The public announcement of ChatGPT by OpenAI on November 23, 2022, marked a significant milestone in the trajectory of AI development, garnering widespread attention and sparking renewed interest in AI applications (Brown et al., 2020). ChatGPT, a state-of-the-art language model, demonstrated remarkable capabilities in generating human-like text and engaging in natural language conversations, elevating the discourse surrounding AI-driven language processing and interaction. This breakthrough, coupled with the proliferation of other AI-powered applications such as Gemini (ex-Bard), Bing, and others, has catalyzed a paradigm shift in how society perceives and engages with AI technologies.

The democratization of AI tools and resources, paired with increased accessibility and user-friendliness, has empowered individuals and organizations to harness the transformative potential of AI in innovative ways, fostering a culture of experimentation and exploration in the realm of AI.

In contemporary discourse, apprehensions regarding the proliferation of AI pervaded the public consciousness, underscoring a spectrum of concerns ranging from its potential threats to traditional employment paradigms (World Bank Group, 2019) to existential risks (Bostrom, 2016) that transcend present socio-economic considerations.
The specter of job displacement looms large as automation and machine learning algorithms encroach upon tasks traditionally performed by humans across diverse industries (Frey and Osborne, 2017).

AI’s uncontrolled proliferation poses existential risks to humanity, extending beyond socio-economic implications. These risks include the potential for technological singularity (Kurzweil, 2024) and the emergence of superintelligent entities with capabilities surpassing human comprehension (Bostrom, 2016).

In combination with these technological advancements, the increase in life expectancy has contributed to significant demographic shifts worldwide (Harper, 2006).

\section{The demographic changes}
\subsection{The demographic changes in Europe}
\begin{figure}[ht]
    \centering
    \includegraphics[width=0.6\textwidth]{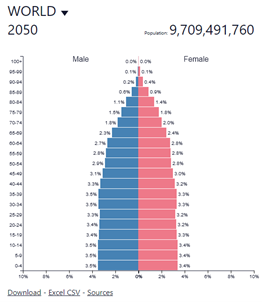}
    \caption{Global Population Projections}
    \label{fig:figure1}
\end{figure}

With advancements in healthcare and an increase in life expectancy, the global population is undergoing significant demographic changes (Harper, 2006). These changes are characterized by a growing number of elderly individuals, particularly those aged 60 years and above. By 2050, it is projected that there will be more people globally over 50 than under 15 (Harper, 2006). See Figure \ref{fig:figure1}. Both Europe and North America are projected to reach their peak population size and commence a demographic decline by the late 2030s (United Nations, 2022).

By 2050, there will be more elderly people than ever before; the number of individuals aged 60 or older is projected to reach approximately 2 billion, accounting for approximately 22 percent of the world’s population (Harper \& Leeson, 2008). The number of individuals aged 80 and above is expected to rise from 90 million to over 400 million, representing approximately 4 percent of the global population (Harper \& Leeson, 2008).

Without immigration, Europe is projected to experience a shortage of 44 million workers by 2050 (Gaines, 2021). Experts estimate that Europe needs to integrate 2-3 million immigrants annually to maintain the 2015 levels of population and economy (CGDEV, 2021). This is even more pronounced in certain European countries. Estimates of the required number of additional working-age adults per country are 7.0 million for Germany, 3.9 million for France, and 3.6 million for the United Kingdom. It is estimated to be 40.1 million for the entire European Union. Alternative projections for the United Kingdom and the European Union are offered by the Center for Global Development which cites the United Nations to indicate that the required number of working-age individuals by 2050 is 61 million (Kenny, 2022).

\begin{figure}[p]
    \centering
    \begin{subfigure}[b]{0.45\textwidth}
        \centering
        \includegraphics[width=\textwidth]{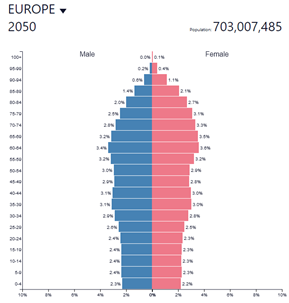}
        \caption{European Population Pyramid in 2050}
        \label{fig:figure2a}
    \end{subfigure}
    \hfill 
    \begin{subfigure}[b]{0.45\textwidth}
        \centering
        \includegraphics[width=\textwidth]{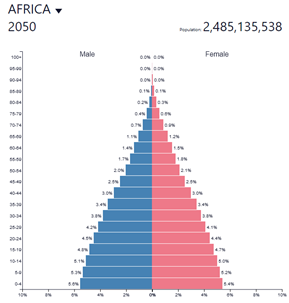}
        \caption{African Population Pyramid}
        \label{fig:figure2b}
    \end{subfigure}
    \par\vspace{0.2cm} 
    \begin{subfigure}[b]{0.45\textwidth}
        \centering
        \includegraphics[width=\textwidth]{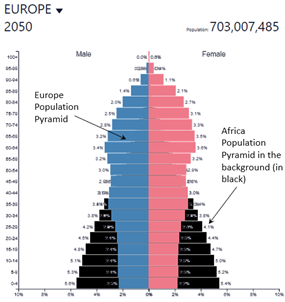}
        \caption{Comparison of European and African Pyramids}
        \label{fig:figure2c}
    \end{subfigure}
    \caption{Population Pyramids for Europe and Africa in 2050}
    \label{fig:population_pyramids}
\end{figure}

The population age pyramid of Europe in 2050 portrays the extent to which the European population is aging (Figure \ref{fig:figure2a}). This is in contrast to Africa’s relatively young population (Figure \ref{fig:figure2b}). This contrast is further emphasized in Figure \ref{fig:figure2c}, where the European pyramid is placed over the African pyramid (in black) to highlight demographic differences, especially in the 0-39 age bracket (PopulationPyramid.net, 2022).

Our interest in the youth inevitably takes us to what is commonly dubbed the “young continent” (Gates, 2018; Mertule, 2015), where 60 percent of the population is under 25. Europe is aging and Africa is young (see Figure \ref{fig:population_pyramids}).

\subsection{The demographic changes in Africa}
Africa has a population of about 1.46 billion people making it home to 18.2 percent of the world’s population (Worldometer, 2024) and the second most populous continent in the world, trailing only Asia. In contrast to all other regions, the African population is expected to increase. Based on simulations run by the United Nations, Africa’s population will rise to 2.5 billion by 2050 (Pison, 2017) when the world population will reach 9.8 billion (Hajjar, 2020) thus making Africans represent more than 25 percent of the world population. Sub-Saharan African countries could account for more than half of the global population growth between 2019 and 2050, and the region’s population is projected to continue to grow through the end of the century (United Nations, 2019).

With 62 percent of its population under 25 in 2019, sub-Saharan Africa has the youngest population in the world (United Nations, 2019), which earned it to also be dubbed the “children’s continent” in Davos (Hajjar, 2020). Other projections suggest that sub-Saharan African youth will continue to grow representing almost 30 percent of the world’s youth by 2050 (United Nations, 2020).

By 2100, one in three people on the planet will be from sub-Saharan Africa, and Nigeria's population will surpass that of China and rank second in the world, just behind India.

By 2050, the Democratic Republic of the Congo (RDC) is projected to reach a population of 200 million, with 30 million residing in the Kinshasa agglomeration alone. Abidjan is anticipated to house 10 million inhabitants, while the collective population of four Sahelian countries\footnote{Namely Mauritania, Mali, Niger, and Chad.} will triple. This growth is exceptional in the annals of human history compared with other regions of the world (Soudan, 2021).

Africa’s youth population is not only growing rapidly, but it is also becoming better educated. As the labor market grows an estimated 12 million new people join the labor force annually (AfDB/OECD/UNDP, 2017). This should be considered in conjunction with the changing educational levels of sub-Saharan African emigration candidates over the past fifteen years. Those come from countries like Côte d’Ivoire, Ghana, or Nigeria (Soudan, 2021).

The changing demographics in Africa, specifically the increasing youth population, will profoundly impact the continent’s future social and economic development (United Nations, 2017).

Yet Africa was found to have the lowest readiness index\footnote{An index set to measure a region's capabilities to equitably use, adopt, and adapt a set number of technologies, mostly information and communication technologies.} scores as developed by the United Nations Conference on Trade and Development to assess national capabilities to equitably use, adopt, and adapt information and communication technologies (ICT) with the least-ready countries being in sub-Saharan Africa (UNCTAD, 2021).

\section{Africa, ICT, and AI}
ICT is Africa’s Achilles’ heel. Sixty percent of youth in sub-Saharan Africa are not online, compared with just 4 percent in Europe (Vincent-Lancrin et al., 2022). Data from the International Telecommunication Union indicate that the percentage of individuals using the Internet in Africa is 37 percent (ITU, 2023), up from 33 percent in 2021 and 28.6 percent in 2019 (ITU, 2021). It is the lowest in the world with an average of 67 percent (ITU, 2023) up from 63 percent in 2021 and 51.4 percent in 2019 (ITU, 2021).

In other regions of the world, developing countries such as Brazil, India, and Mexico have seen higher growth in digital adoption six months before April 2021 than in developed countries between 2020 and 2021 (Hajro et al., 2021). The COVID-19 epidemic has accelerated digitalization and produced an advantageous climate (McKinsey \& Co., 2022).

Between 2020 and 2021, the number of tech start-ups in Africa tripled to around 5,200 companies. According to McKinsey \& Co. (2022), fintech companies make up approximately half of these. However, some regions face more challenges than others in initiating this process because of the absence of adequate bandwidth, Internet connections, and other technologies.

African countries such as South Africa, Kenya, Egypt, and Nigeria are already leading in digital transformation. Despite having relatively low rates of banking penetration, certain African nations have made significant strides in the areas of digital payments and financial inclusion even before the outbreak of the pandemic, with rates of around 15 percent. Cash is still used in over 90 percent of transactions, therefore there is a huge development opportunity for fintech revenue (McKinsey \& Co., 2022).

ICTs have this unique characteristic of allowing leapfrogging (Ojanperä et al., 2016), which means that in their development countries and societies can leap across several generations and stages of technology development. This has also been observed in Africa (McKinsey \& Co., 2022).

One of the first examples of such was seen in Latin America where around 46 percent of households had no fixed-line service in 2003 and mobile telephony was adopted even faster and easier precisely because land-line penetration was very low, particularly in poor, rural regions\footnote{In 2003 Nicaragua had just 180,000 fixed lines for 5.5 million people. Experts estimate that the number of mobile lines there will grow at an annual rate of 9 percent through 2008, compared with just 4 percent a year for fixed-line service. See Smith (2003).}.

The state-of-the-art Frontier Technologies (UNCTAD, 2021)\footnote{These include the Internet of Things, Concentrated Solar Power, Blockchain, Nanotechnology, Big Data, 5G, Biofuels, Electric Vehicles, Gene Editing, Robotics, Drone Technology, 3D Printing, Wind Energy, Biogas and biomass, Green Hydrogen, Solar photovoltaic, and AI, of which eight are directly related to ICT, labeled Industry 4.0 technologies in UNCTAD (2021; 2023).} in Africa are evolving rapidly. A review of AI research in finance in Africa indicates increasing interest in and exploration of Frontier Technologies such as AI (Gikunda, 2023).

As we pointed out earlier, AI has gained significant traction globally, with both opportunities and challenges arising from its widespread adoption (Kalyanakrishnan et al., 2018). AI has shown success in various fields, including healthcare, medicine, public services, and education (Kusters et al., 2020). However, there are disparities in the adoption and development of AI between countries and regions (Okolo et al., 2023).

AI is a potential solution to many of the challenges faced by the continent, including unemployment and poor infrastructure (Longbing, 2021). However, sub-Saharan Africa also exhibited the lowest index among all the geographical groups. Indeed, there are still significant barriers to the catch-up of Frontier Technologies in Africa (Ezeani, 2022; UNCTAD, 2021).

Although some African countries have started exploring AI applications, overall progress and investment in AI in the continent lag behind global leaders such as China and the United States (Gikunda, 2023).

Several challenges hinder the effective adoption and implementation of AI in Africa; including limited digital literacy, inadequate infrastructure, and insufficient government support and regulation.

Additionally, the lack of AI research in Africa contributes to the barriers faced in implementing AI technology. While the share of global AI-related peer-reviewed journal publications in 2020 was 13.3 percent for Europe and Central Asia, it was 0.3 percent for sub-Saharan Africa (Zhang et al., 2021). Two years later, it was 17.20 percent in Europe and Central Asia and 0.77 percent in sub-Saharan Africa (Maslej et al., 2023).

\begin{figure}[ht]
    \centering
    \includegraphics[width=0.9\textwidth]{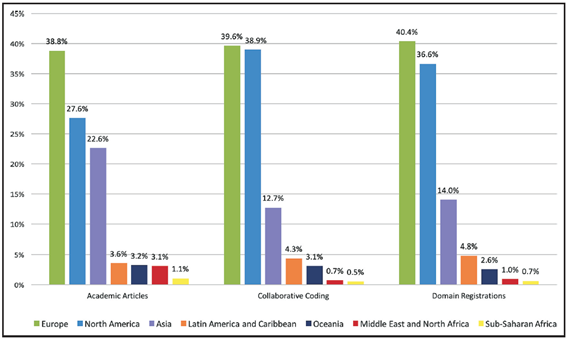}
    \caption{AI Adoption in Africa}
    \label{fig:image3}
\end{figure}

While Europe’s contribution to academic articles amounts to 38.8 percent, sub-Saharan Africa barely reaches 1.1 percent. When Europe’s contribution to domain registration was 40.4 percent, that of sub-Saharan Africa barely reached 0.7 percent (see Figure \ref{fig:image3}, Ojanperä et al., 2017).

In the UNCTAD (2021, 2023) Frontier Technologies Big Data stands tall among so-called Industry 4.0 technologies; data go hand in hand with AI, and there is little AI without data and datasets, both pillars of a digital culture (LeCun et al., 2015).

Many Africans do not have the skills to effectively create and use data. The 2023 AI-Index Report shows that India has the highest relative penetration of AI skills with a score of 3.23 percent in 2022. It is worth noting that despite South Africa being ranked 13th in 2019 (Zhang et al., 2021), no African nation made it to the top 15 in 2022 (Maslej et al., 2023). In summary, although there is global recognition of the potential benefits of AI, its adoption and implementation in Africa are still in their early stages (Okolo et al., 2023; Kusters et al., 2020).

\section{The lack of a digital culture in Africa}
Unfortunately, sub-Saharan Africa has also been reputed to have comparatively weak national statistical institutions (Ojanperä et al., 2016; Van Belle, 2018). On a global scale, Africa has the lowest average level of statistical capacity. Merely half of the African countries have conducted more than two comparable household surveys within the last decade (van Belle et al., 2017). Additionally, only 29 percent of African nations have published household surveys containing educational data since 2005 (Van Belle et al., 2017).

In 2023, Europe contributed 19.3 percent and the Middle East and Africa contributed 6.8 percent to global data creation. The Middle East and Africa are projected to have the fastest growth, with their contribution rising to 9.6 percent by 2027 (IDC, 2023).

The gap in data availability is mainly caused by various challenges such as limited access to a significant amount of data, poor data quality, inadequate storage infrastructure, lack of regulatory policies for data (Abebe et al., 2021), and insufficient knowledge and skills in data management (Gwagwa et al., 2020). There is also an issue with an underlying digital culture which is lacking. Some African governments' imposition of taxes on digital content creation (Mpofu \& Moloi, 2022) is not conducive to developing a culture of data creation. Such initiatives will tend to hinder the continent's ability to enhance its contribution to global data generation (IDC, 2023).

Additionally, most existing datasets in Africa are still paper-based, and organizations are often unwilling to digitize them. As a result, Africa lacks the necessary African-origin datasets that are relevant, reliable, and of high quality to support AI research and innovation (Sinde et al., 2023) and even to foster and perpetuate an existing culture.

There is a significant difference in how AI applications work in the African context compared with other regions. This disparity has limited the effectiveness of AI tools and has even worsened inequities in Africa (Okolo et al., 2023). Unfortunately, this makes it difficult for the local workforce to develop AI skills and expertise. And Africa’s significant gap in AI data hinders the continent’s progress in research, development, and innovation (Kiemde \& Kora, 2020; Gwagwa et al., 2020).

\section{The future of work in the context of AI}
Goldman Sachs estimated that over ten years, more jobs in developed countries will be exposed to automation than in emerging countries (Hatzius et al., 2023). Conversely, according to Accenture and Frontier Economics it is in the developed countries that AI will cause labor productivity to continue to increase by up to 40 percent until 2035 (World Economic Forum, 2020).

The World Economic Forum's Future of Jobs Report released in 2023 (World Economic Forum, 2023), anticipates that over the next five years (2023-2027), approximately 83 million jobs will disappear, while 69 million new positions will emerge, resulting in a net decrease of 14 million jobs, which is equivalent to 2 percent of current employment.

McKinsey \& Co. suggested that as many as half of the tasks performed in current jobs could be automated between 2030 and 2060, with the midpoint around 2045. This projection indicates an acceleration of approximately ten years compared to their earlier estimations. However, the pace of workforce transformation and automation adoption is expected to be quicker in developed nations and slower in lower-wage countries (McKinsey \& Co., 2023).

While the World Economic Forum is not explicit as the cause of the loss of jobs and the share of AI’s responsibility in that, we can already posit, as McKinsey reports, that in the United States and Europe, two-thirds of jobs face some level of AI automation. According to Goldman Sachs, 18 percent of the work can be automated (Hatzius et al., 2023; McKinsey \& Co, 2023). Based on various sources (OECD, IMF, etc.), Vocelka computed that between 2019 and 2050, the average number of hours “worked” by IA will increase from 0 to 333, whereas the average number of hours worked by humans will drop from 1,749 to 1,421, paralleling productivity gains from 15 percent to 81 percent over the same period (Vocelka, 2023).

Despite this, sub-Saharan Africa must create 20 million jobs each year for 20 years to keep up with population growth, as stated in a 2018 report by the International Monetary Fund (Abdychev et al., 2018). These jobs must be in higher-valued, higher-skilled occupations. However, the creation of 400 million jobs is daunting. Incidentally, Africa requires more jobs whereas Europe requires more workers.

It is important to approach assumptions about the impact of AI on labor with caution. This is because the composition of the labor force in Africa differs significantly from that in Europe, as do job requirements. Therefore, AI’s effects on employment are expected to vary considerably.

Furthermore, it is difficult to predict which skills and labor force will be required in the future because of the unpredictable nature of AI. However, there is no certainty of its long-term impact. AI is advancing so rapidly that even industry experts struggle to track, let alone predict its full implications and applications across diverse domains.

\section{Conclusion}
Africa's changing demographics, especially the rising youth population, will have a significant impact on the continent's future social and economic development (PRB, 2019).

By 2050 Africa will be the most populous and youngest continent in the world but will remain one of the least digital overall; low-skilled African migrants will be limited to low-wage employment. But the continuously widening digital gap between sub-Saharan African nations and developed regions of the world will impede efforts to further diversify their economies, create jobs, and reduce economic dependence (UNCTAD, 2021).

Nonetheless, it is important to consider possible scenarios for the evolution of Africa over the next 25 years.

1. The best-case scenario for Africa would be to catch up technologically with the rest of the world through leapfrogging for example (Ojanperä et al., 2016). This would enable the continent to level up with the West and share similar technological challenges. Additionally, this could lead to African immigrants occupying jobs that have not yet been replaced by AI, thereby contributing significantly to the European economy.

2. If Africa fails to catch up technologically, it could face a bleak future as predicted by UNCTAD, which states that "China and the United States will reap the highest economic benefits from AI, while Africa and Latin America are likely to experience the smallest gains" (UNCTAD, 2019, p. 8). This, combined with impressive demographic changes will lead to an extremely difficult landscape.

In Africa, governments face greater challenges than in Europe because of youth bulge with both opportunities and challenges. On one hand, the growing youth population can serve as a valuable asset for driving economic growth and innovation. However, if not properly addressed, it can lead to high youth unemployment rates and social unrest (United Nations, 2017). As stated by the authors of the State of the World Population in the 2023 report, “Too many young people? Destabilizing. Too many old people? A burden. Too many migrants? A threat” (UNFPA, 2023; p. 6).

Therefore it is important to consider the impact of AI on employment in the next two decades. Have the predictions put forth earlier about job needs in Europe considered the effects of AI on employment? The answer to this question should also consider that robots and new biological beings are plausible extensions of current technological developments.

In this paper, we suggest that African youth will play a significant role in shaping the future of the European economy because Africa is poised to become a vital provider of workers to Europe. Looking further into the future, the possibility also exists that today’s African diaspora will be tomorrow’s returnees, who will by ricochet play a significant role in shaping the future of the African economy too.

Having acquired skills in Europe, returnees will contribute greatly to economic (and technological) development in their home countries. In the 1970s, diaspora returnees played a significant role in contributing to the economic boost experienced by South Korea (Bahar, 2020) and China (Cheung, 2004). Individuals who had emigrated from their home countries and then returned with skills, knowledge, and networks acquired abroad, have brought valuable expertise and entrepreneurship (Bahar, 2020; Lin, 2010) changing brain drain to brain gain (Chrysostome \& Nkongolo-Bakenda, 2019).

African governments and policymakers must not sleep on their laurels. They should not only continue to prioritize investments in education, skill training, and job creation to harness the potential of the youth population but also overcome the lack of digital skills and cultivate a digital culture to fully realize the potential of the data economy.

\section*{References}
Abebe, R., Aruleba, K., Birhane, A., Kingsley, S., Obaido, G., Remy, S. L., and Sadagopan, S. (2021). Narratives and counternarratives on data sharing in Africa. \emph{Proceedings of the 2021 ACM Conference on Fairness, Accountability, and Transparency}, 329–341. \url{https://doi.org/10.1145/3442188.3445897} \\

Abdychev, A., Alonso, C., Alper, E., Desruelle, D., Kothari, S., Liu, Y., Perinet, M., Rehman, S., Schimmelpfennig, A., and Sharma, P. (2018). \emph{The future of jobs in sub-Saharan Africa}. International Monetary Fund. \\

African Development Bank, Organisation for Economic Co-operation and Development, and United Nations Development Programme. (2017). \emph{African economic outlook 2017}. OECD Publishing. \url{https://doi.org/10.1787/aeo-2017-en} \\

Bahar, D. (2020). Diasporas and economic development: A review of the evidence and policy. \emph{Comparative Economic Studies, 62}(2), 200–214. \url{https://doi.org/10.1057/s41294-020-00117-0} \\

Bostrom, N. (2016). \emph{Superintelligence: Paths, dangers, strategies}. Oxford University Press. \\

Brown, T. B., Mann, B., Ryder, N., Subbiah, M., Kaplan, J., Dhariwal, P., Neelakantan, A., Shyam, P., Sastry, G., Askell, A., Agarwal, S., Herbert-Voss, A., Krueger, G., Henighan, T.,
Child, R., Ramesh, A., Ziegler, D. M., Wu, J., Winter, C., ... and Amodei, D. (2020). Language models are few-shot learners. \emph{Advances in Neural Information Processing Systems, 33}, 1877–1901. \url{https://arxiv.org/abs/2005.14165} \\

Center for Global Development. (2021). \emph{Migration and demography: Europe's aging workforce}. \url{https://www.cgdev.org/article/europe-be-short-44-million-workers-2050-without-increased-immigration-new-study-finds} \\

Cheung, G. C. K. (2004). Chinese diaspora as a virtual nation: Interactive roles between economic and social capital. \emph{Political Studies, 52}(4), 664–684. \url{https://doi.org/10.1111/j.1467-9248.2004.00501.x} \\

Chrysostome, E. V., and Nkongolo-Bakenda, J.-M. (2019). Diaspora entrepreneurship: A promising driver for economic development in Africa. \emph{Journal of African Business, 20}(4), 441–457. \url{https://doi.org/10.1080/15228916.2019.1641306} \\

Frey, C. B., and Osborne, M. A. (2017). The future of employment: How susceptible are jobs to computerisation? \emph{Technological Forecasting and Social Change, 114}, 254–280. \url{https://doi.org/10.1016/j.techfore.2016.08.019} \\

Gaines, J. (2021). Europe’s aging population and workforce shortage. \emph{Economic Review}, 12(3), 45–60. \\

Gates, B. (2018). \emph{What you need to know about the world's youngest continent}. GatesNotes. \url{https://www.gatesnotes.com/Africa-the-Youngest-Continent} \\

Gikunda, K. (2023). \emph{Empowering Africa: An in-depth exploration of the adoption of artificial intelligence across the continent}. \url{https://arxiv.org/abs/2401.09457} \\

Gwagwa, A., Kachidza, E., and Kazim, E. (2020). Data governance in Africa: Challenges and opportunities. \emph{AI \& Society, 35}(4), 1027–1036. \url{https://doi.org/10.1007/s00146-020-01058-7} \\

Hajjar, H. (2020). \emph{Africa on the move: The continent of youth}. Migration Policy Institute. \url{https://www.migrationpolicy.org/article/africa-move-continent-youth} \\

Hajro, M., Bandiera, O., and Vitanov, S. (2021). \emph{Digital adoption during COVID-19}. World Bank Blogs. \url{https://blogs.worldbank.org/opendata/digital-adoption-during-covid-19} \\

Harper, S. (2006). Mature societies: Planning for our future selves. \emph{Daedalus, 135}(1), 20–31. \url{https://doi.org/10.1162/daed.2006.135.1.20} \\

Harper, S., and Leeson, G. W. (2008). Ageing populations: Problems and solutions. \emph{The World Economy, 31}(6), 737–762. \url{https://doi.org/10.1111/j.1467-9701.2008.01072.x} \\

Hatzius, J., Briggs, J., Kodnani, D., and Pierdomenico, G. (2023). \emph{The potentially large effects of artificial intelligence on economic growth}. Goldman Sachs. \url{https://www.ansa.it/documents/1680080409454_ert.pdf} \\

International Data Corp. (2023). \emph{Worldwide Global DataSphere and Global StorageSphere structured and unstructured data forecast, 2023–2027} (Document No. US50397723). \\

International Telecommunication Union. (2021). \emph{Measuring digital development: Facts and figures 2019}. \url{https://www.itu.int/en/ITU-D/Statistics/Documents/facts/FactsFigures2021.pdf} \\

International Telecommunication Union. (2023). \emph{Measuring digital development: Facts and figures 2023}. \url{https://www.itu.int/dms_pub/itu-d/opb/ind/d-ind-ict_mdd-2023-1-pdf-e.pdf} \\

Kalyanakrishnan, S., Panicker, R. A., Natarajan, S., and Shreya, R. (2018). Opportunities and challenges for artificial intelligence in India. In \emph{Proceedings of the 2018 AAAI/ACM Conference on AI, Ethics, and Society} (pp. 164–170). \url{https://doi.org/10.1145/3278721.3278738} \\

Kenny, C. (2022). \emph{Good news: Africa needs more jobs while Europe needs more workers}. Center for Global Development. \url{https://www.cgdev.org/blog/good-news-africa-needs-more-jobs-while-europe-needs-more-workers} \\

Kiemde, S. M. A., and Kora, A. D. (2020). The challenges facing the development of AI in Africa. \emph{IEEE International Conference on Advanced Technologies and Management Research in ICT}, 1–6. \url{https://doi.org/10.1109/icatmri51801.2020.9398454} \\

Kurzweil, R. (2024). \emph{The singularity is near: When humans transcend biology}. Penguin Publishing Group. \\

Kusters, R., Mishra, V., and Singh, A. (2020). AI applications in healthcare and education in developing countries. \emph{Journal of Artificial Intelligence Research, 12}, 45–60. \\

LeCun, Y., Bengio, Y., and Hinton, G. (2015). Deep learning. \emph{Nature, 521}(7553), 436–444. \url{https://doi.org/10.1038/nature14539} \\

Lin, X. (2010). The diaspora solution to economic development: Lessons from China. \emph{Journal of International Migration and Integration, 11}(3), 285–304. \url{https://doi.org/10.1007/s12134-010-0140-2} \\

Longbing, C. (2021). AI for social good: Opportunities in Africa. \emph{IEEE Intelligent Systems, 36}(4), 3–7. \\

Maslej, N., Fattorini, L., Brynjolfsson, E., Etchemendy, J., Ligett, K., Lyons, T., Manyika, J., Ngo, H., Niebles, J. C., Parikh, R., and Shoham, Y. (2023). \emph{AI index report 2023}. Stanford University. \\

McKinsey \& Co. (2022). \emph{Fintech in Africa: The end of the beginning}. \url{https://www.mckinsey.com/~/media/mckinsey/industries/financial%20services/our%20insights/fintech%20in%20africa%20the%20end%20of%20the%20beginning/fintech-in-africa-the-end-of-the-beginning-f.pdf} \\

McKinsey \& Co. (2023). \emph{The economic potential of generative AI: The next productivity frontier}. \url{https://www.mckinsey.com/capabilities/mckinsey-digital/our-insights/the-economic-potential-of-generative-ai-the-next-productivity-frontier} \\

Mertule, D. (2015). Africa: The young continent. \emph{Global Demographics Review, 8}(2), 12–18. \\

Mpofu, F. Y., and Moloi, T. (2022). Digital taxation and its impact on content creation in Africa. \emph{African Journal of Science, Technology, Innovation and Development, 14}(3), 789–801. \url{https://doi.org/10.1080/20421338.2021.1908440} \\

Ojanperä, S., Graham, M., and Zook, M. (2016). Leapfrogging in ICT: Evidence from Africa. \emph{Development Policy Review, 34}(5), 645–667. \url{https://doi.org/10.1111/dpr.12181} \\

Ojanperä, S., Graham, M., and Zook, M. (2017). Mapping global AI research contributions. \emph{Journal of Digital Geography, 3}(1), 22–35. \\

Okolo, C. T., Kamara, I., and Dell, N. (2023). AI in Africa: Contextual challenges and opportunities. \emph{Proceedings of the ACM on Human-Computer Interaction, 7}(CSCW1), 1–20. \url{https://doi.org/10.1145/3579478} \\

Pison, G. (2017). Population projections for Africa: 2050 and beyond. \emph{Population Studies, 71}(3), 295–310. \url{https://doi.org/10.1080/00324728.2017.1351956} \\

Population Reference Bureau. (2019). \emph{Africa’s future: Youth and the data defining their lives}. \url{https://www.prb.org/wp-content/uploads/2019/10/Status-of-African-Youth-SPEC.pdf} \\

Sinde, R., Diwani, S. A., Leo, J., Kaijage, S., and Nyoni, T. (2023). AI for Anglophone Africa: Unlocking its adoption for responsible solutions in academia-private sector. \emph{Frontiers in Artificial Intelligence, 6}, 1133677. \url{https://doi.org/10.3389/frai.2023.1133677} \\

Soudan, F. (2021). \emph{Africa’s demographic explosion: Opportunities and challenges}. Le Monde Afrique. \\

United Nations. (2017). \emph{World population prospects: The 2017 revision}. United Nations Department of Economic and Social Affairs. \\

United Nations. (2019). \emph{World population prospects 2019}. United Nations Department of Economic and Social Affairs. \\

United Nations. (2020). \emph{Youth population trends in sub-Saharan Africa}. United Nations Population Division. \\

United Nations. (2022). \emph{World population prospects 2022}. United Nations Department of Economic and Social Affairs. \\

United Nations Conference on Trade and Development. (2019). \emph{Digital economy report 2019}. United Nations. \\

United Nations Conference on Trade and Development. (2021). \emph{Technology and innovation report 2021}. United Nations. \\

United Nations Population Fund. (2023). \emph{State of world population 2023}. UNFPA. \\

Van Belle, J.-P. (2018). Statistical capacity in sub-Saharan Africa: Challenges and solutions. \emph{African Statistical Journal, 19}, 45–60. \\

Van Belle, J.-P., Smith, R., and Jones, T. (2017). Assessing statistical capacity in Africa. \emph{Journal of Development Data, 5}(2), 33–48. \\

Vincent-Lancrin, S., Hattingh, D., and Jacotin, G. (2022). \emph{Digital divides in education: A global perspective}. OECD Publishing. \\

Vocelka, A. (2023). AI governance for a prosperous future. In R. Schmidpeter and R. Altenburger (Eds.), \emph{Responsible artificial intelligence} (pp. 17–90). Springer. \\

World Bank Group. (2019). \emph{The changing nature of work} (World Development Report). \url{http://documents.worldbank.org/curated/en/816281518818814423/pdf/2019-WDR-Report.pdf} \\

World Economic Forum. (2020). \emph{The future of jobs: AI and productivity}. World Economic Forum. \\

World Economic Forum. (2023). \emph{Future of jobs report 2023}. \url{https://www3.weforum.org/docs/WEF_Future_of_Jobs_2023.pdf} \\

Worldometer. (2024). \emph{Current world population}. \url{https://www.worldometers.info/world-population/#region} \\

Zhang, D., Mishra, S., Brynjolfsson, E., Etchemendy, J., Ganguli, D., Grosz, B., Lyons, T., Manyika, J., Niebles, J. C., Sellitto, M., Shoham, Y., Clark, J., and Perrault, R. (2021). \emph{AI index report 2021}. Stanford University.
\end{document}